\documentclass[preprint,prb,aps,showpacs]{elsart}
\usepackage{graphics,subfigure,epsfig,amssymb}
\graphicspath{{./figs/}}   
\newcommand{\qav}[1]{\left \langle #1 \right \rangle}     
\newcommand{\pa}[1]{\left ( #1 \right )}      
\newcommand{\ab}[1]{\left | #1 \right |}     
\newcommand{\ev}[1]{\langle #1 \rangle}  
\newcommand{\mrm}[1]{\mathrm{#1}}  
\begin{document}   
\begin{frontmatter}
\title{Microscopic modeling of photoluminescence of 
strongly disordered semiconductors\thanksref{acknowledgments}}   

\thanks[acknowledgments]{
The authors are thankful to R.\ Zimmermann and S.\ D.\ Baranovskii for
valuable discussions. This work has been supported by the
Optodynamic Center of the Philipps-University Marburg, and by the
Deutsche Forschungsgemeinschaft (DFG) through the Quantum Optics in
Semiconductors Research Group. T.\ M.\ thanks the DFG for support via
a Heisenberg fellowship (ME 1916/1). I.\ V.\ thanks for financial
support from the OTKA (Hungarian Research Fund) under contract
Nos. T042981 and T046303.
We thank the John von Neumann Institut f\"ur Computing (NIC),
Forschungszentrum J\"ulich, Germany, for grants for extended CPU time
on their supercomputer systems.
}

\author{P. Bozsoki\corauthref{cor}}, 
\corauth[cor]{Corresponding author.}\ead{peter.bozsoki@physik.uni-marburg.de}
\author{M. Kira},
\author{W. Hoyer},
\author{T. Meier},
\author[BP]{I. Varga},
\author{P. Thomas},
\author{S.W. Koch}
\address{Department of Physics and Material Sciences Center,
Philipps University, Renthof 5, D-35032 Marburg, Germany}
\address[BP]{Elm\'eleti Fizika Tansz\'ek,
Fizikai Int\'ezet, Budapesti M\H{u}szaki \'es
Gazdas\'agtudom\'anyi Egyetem, Budafoki \'ut 8, H-1111 Budapest, Hungary}

\date{\today}   
  
\begin{abstract}   
A microscopic theory for the luminescence of ordered semiconductors 
is modified to describe photoluminescence of strongly disordered  
semiconductors. 
The approach includes both
diagonal disorder and the many-body Coulomb interaction. 
As a case study, the light emission of a correlated plasma
is investigated numerically
for a one-dimensional two-band tight-binding model.
The band structure of the underlying ordered system is 
assumed to correspond to either a direct or an indirect semiconductor.   
In particular, luminescence 
and absorption spectra are computed for various levels of disorder  
and sample temperature to determine thermodynamic relations, 
the Stokes shift, and the radiative lifetime distribution.  
\end{abstract}   

\end{frontmatter}
\maketitle   
   

\section{Introduction}   
Photoluminescence experiments are widely used as a diagnostic tool  
to investigate the electronic structure of
semiconductors and semiconductor heterostructures in particular,
also in the presence of disorder. 
Such measurements can, e.g., provide information about   
the electron-hole many-body states related to the dynamics of  
relaxation of the optically generated excitations. 
In particular, temperature-dependent spectra can be used to  
study localization of the optical excitations
\cite{wir,zim,Daly95,Davey88,Skolnick86}.
Time-resolved  
data and radiative lifetime distributions measured by   
frequency-response-spectroscopy \cite{Fuhs} give further  
information about the electronic states and relevant processes  
in disordered semiconductors.  
  
In order to extract meaningful information from the luminescence data, 
a theoretical model which incorporates the relevant physical processes
is required. Starting from a model which is able to
reproduce the main features of the experimental data
for reasonable parameters, one tries to   
arrive at a possible interpretation.   
So far, dynamical aspects as well as spectral signature of
the photoluminescence of disordered semiconductors have usually been
computed using Monte-Carlo approaches 
\cite{wir,Gruening04,Zukauskas03,Don04,pairs,Rubel05,Boulitrop83},
The underlying model is either based on excitonic excitations 
\cite{wir,Gruening04,Zukauskas03,Don04} or, alternatively,   
independent electron and hole excitations are assumed 
\cite{pairs,Rubel05,Boulitrop83}.

The Monte Carlo methods, discussed above, phenomenologically 
describe the following processes: 
(i) The photo-excitation with excess energy creates electron-hole pairs  
in states that are energetically well above the emitting states.   
(ii) Depending   
on the material system under study, fast incoherent relaxation  
of either excitons or independent electrons and holes into energetically  
lower lying states close to the band edges is assumed.  
(iii) As these states are presumably localized in disordered semiconductors  
in the absence of coupling to phonons,  
a slower phonon-induced relaxation among these states takes place which is 
governed by their mutual distance, the localization  
length, the energetic density of states, and the temperature. 
(iv) These relaxation processes compete with radiative   
and non-radiative recombination. In the exciton picture, 
the radiative recombination rate is given by the intrinsic  
structure, i.e., by the optical matrix elements of the correlated 
electron-hole pair  
\cite{wir}.
For independent electrons and holes it is their mutual  
distance that determines their recombination rate via a  
radiative tunneling process \cite{pairs}.  

This scenario suggests the following model parameters:
localization length (sometimes allowed to depend  
on energy), total density of localized states in the band tails,  
the form of the density of states in the tail  
region, and an attempt-to-escape frequency determining the  
phonon-induced tunneling (hopping) into energetically higher  
or lower lying states. By choosing a model and  
fitting the simulated spectra to experimental  
ones one finds  
a combination of parameters which represent the data in an optimal way.   
The model and its parameters can be further substantiated if  
one succeeds to reproduce simultaneously data for various temperatures  
and also time resolved spectra.    
 
Although this procedure is widely used and has been successfully applied 
in many cases e.g.~\cite{wir,Gruening04,Zukauskas03,Don04},
it is not very satisfactory from a more fundamental point of view.   
Instead of using a phenomenological description, one should be able 
to analyze the 
luminescence of disordered semiconductors
using a microscopic approach which is 
based on a more realistic model of a disordered structure. 
An attempt into this direction has been presented
by Zimmermann and Runge \cite{zim}
who have used an exciton picture for the case of a 
weakly disordered system. 
An adequate and comprehensive microscopic theory for the luminescence 
in the presence of
disorder should simultaneously
include the phonon-induced relaxation, radiative and  
non-radiative recombination, and the many-particle Coulomb interaction.  
Microscopic approaches have recently been presented 
on the basis of an effective-mass approach considering both
bound-excitonic
and separate electron-hole recombination processes.
\cite{Singh02,singh}.

For the case of perfectly ordered  
systems, a microscopic theory of semiconductor luminescence  
has been developed recently \cite{Kira98,PQE,Kira01,Chatterjee04}.
Although microscopic aspects of the Coulomb and  
electron-phonon interactions have been included in this approach,  
it is by no means trivial to extend this theory 
to disordered structures since one can  
calculate the single-particle eigenstates only  
for relatively small systems. In addition, an adequate 
analysis of the phonon-induced  
relaxation in a disordered system requires the treatment of  
many-acoustic-phonon-induced hopping processes 
in a system of interacting electrons, because energy differences
involved in the relaxation processes in strongly disordered
semiconductors are much larger than the Debye energy.
A fully microscopic theory meeting these challenges does not yet
exist.
 
As a step towards this goal, we present in this paper a treatment of  
luminescence in a one-dimensional tight-binding  
model of a disordered two-band semiconductor. 
We do not intend to model any existing 
bulk semiconductor or semiconductor heterostructure. We rather perform 
a case study, which treats disorder and Coulomb interaction 
on an equal footing. Applying the theory worked out in
\cite{Kira98,PQE,Kira01,Chatterjee04,Hoyer03,Hoyer04} 
for ordered semiconductors we 
arrive at an expression that describes radiative
electron-hole recombination in a
disordered system without the necessity to
separately modelling bound excitonic vs. separate-pair (plasma) processes
(see, e.g.,  \cite{singh} for such an approach). 
The price we have to pay is to completely neglect the
explicit description of phonon-induced relaxation and hopping processes. 
Instead,
a stationary situation which is characterized by
static populations of the states is assumed.
These are modeled by
quasi-equilibrium Fermi-Dirac distributions for electrons and holes,   
implicitly assuming that  
the relaxation and intraband dynamics of carriers is much faster than the  
radiative process. 
It is known that the initial relaxation is fast, i.e., 
on the order of ps~\cite{singh,schwarz}.
In our model, we assume that  
this is also the case for subsequent steps such that  
quasi-stationary thermal single-particle distributions  
are always maintained. 
In the present paper, we do  
not discuss the interesting and important 
issue of exciton formation. 
From previous publications which investigated ordered semiconductors 
\cite{Kira01,Hoyer03,Hoyer04}, 
it is known that: i) exciton formation is relatively slow, 
particularly for temperatures above 30\,K. ii) Even when an appreciable 
amount of excitons is present, the optically active ones recombine rapidly 
leaving behind a correlated plasma as the dominating source of 
luminescence in many 
experimentally-relevant situations.\cite{Chatterjee04} 
iii) In addition, the spectral positions of the dominant 
emission resonances do not  
strongly depend on the microscopic source, i.e., are similar
for excitonic and independent electron and hole populations. 
Thus, our luminescence analysis should be reasonable 
for the case of
quasi-stationary situations in which optically active exciton 
populations can be neglected.
In ordered semiconductors, such conditions are typically  
realized for sufficiently high sample  
temperatures.\cite{Chatterjee04} 
For situations in which exciton populations dominate,
our results are expected to underestimate the strength of the 
excitonic resonance, however, the peak position should be
correctly described.  
 
The semiconductor model is based on an ordered
system that can either be chosen to represent semiconductors with
a direct or an indirect band gap.
We can, therefore, study to  
what extent, as function of the strength of the   
disorder, the direct or an indirect nature 
of the model influences the luminescence  
and the absorption spectra. 
This allows us to discuss in detail the  
structure of the luminescence and absorption spectra,  
their thermodynamic relation, 
the dependence on the model parameters, 
the Stokes shift, and the  
radiative lifetime distribution for emission from 
an electron-hole plasma. 
  
This paper is organized as follows: 
The model and the theoretical  
approach are described in Section~\ref{model} 
and Section~\ref{approach},
respectively.
The results are presented and  
discussed in Section~\ref{results}. 
This work ends with a 
discussion in a concluding 
Section~\ref{con} that also presents an 
outlook for further work.
 
\section{The model}   
\label{model}  
 
We consider a one-dimensional tight-binding model 
of a disordered two-band semiconductor.
The total Hamiltonian is    
\begin{equation}
\widehat{H}_{\mrm{total}} = \widehat H_0 +   
\widehat H_{\mrm{Cb}} + \widehat  H_{\mrm{EM}} . 
\label{eq_ham}  
\end{equation}
Here, $\widehat H_0$ is the single-particle  
Hamiltonian describing  
the valence ($h$) and the conduction ($e$) band,  
\begin{equation}      
  \widehat H_0 = \sum_{\alpha}   
\epsilon_{\alpha}^e \widehat e^{\dag}_{\alpha} \widehat e_{\alpha} +   
  \sum_{\beta} \epsilon_{\beta}^h    
\widehat h^{\dag}_{\beta} \widehat h_{\beta} ,  
\end{equation}   
where $\widehat e^{\dag}_{\alpha}$ ($\widehat e_{\alpha}$) and     
$\widehat h^{\dag}_{\beta}$ ($\widehat h_{\beta}$) denote  
electron and hole creation (annihilation) operators,  
respectively. 
In Eq.~(\ref{eq_ham}), the Coulomb interaction among carriers is described by 
$\widehat H_{\mrm{Cb}}$ and
$\widehat   
H_{\mrm{EM}}$ represents the Hamiltonian of the electromagnetic field  
and its interaction with the carrier system.  
  
\begin{figure}[htb!]\centering  
\resizebox{18cm}{!}{\includegraphics{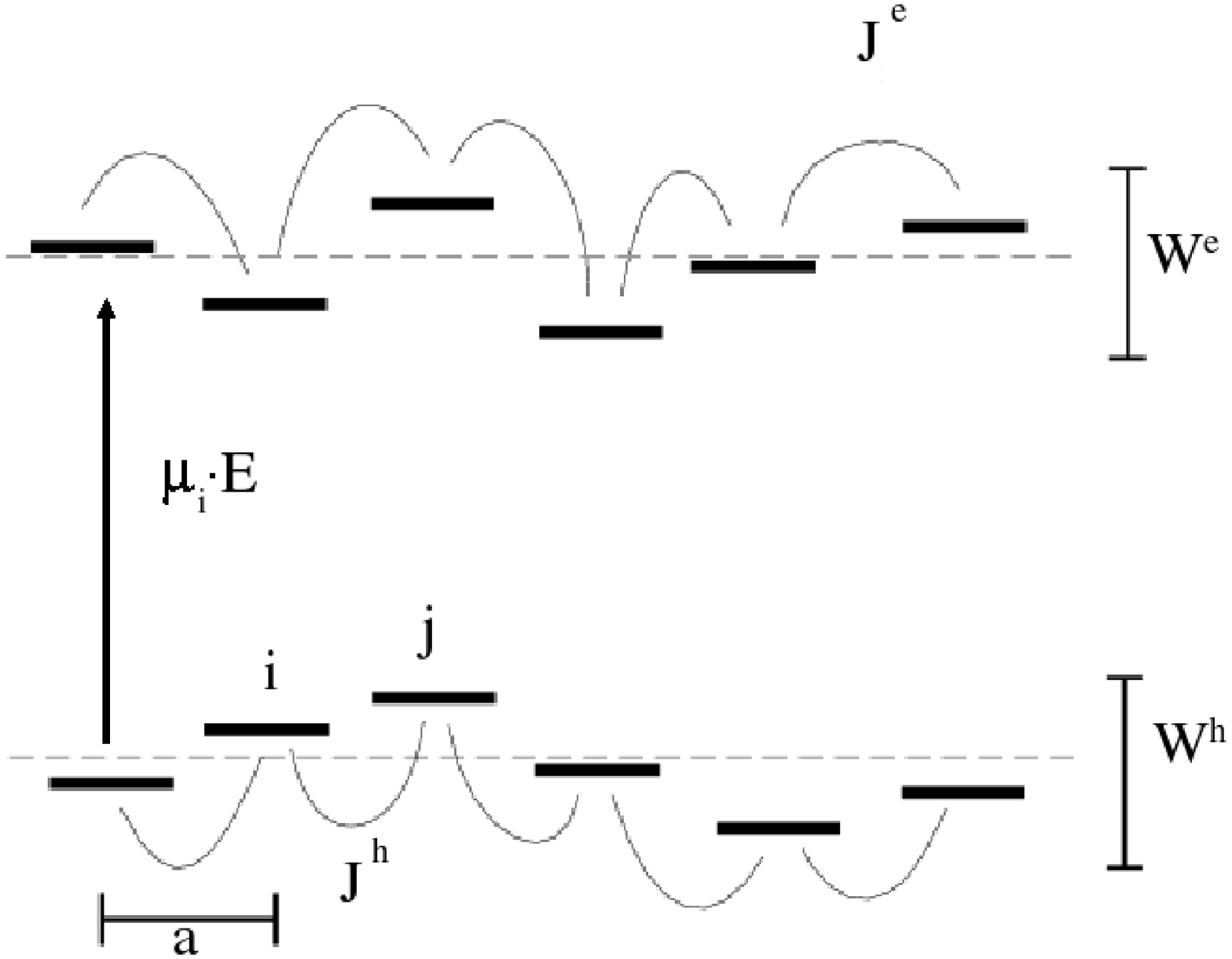}}   
\caption{\small Pictorial representation of the tight-binding model.  
The energies of the isolated sites are distributed over a disorder  
width $W^e$ and $W^h$ and coupled to the nearest neighbors  
via couplings $J^e$ and $J^h$. The local excitation at site  
$i$ with matrix element $\mu_i$ generates a polarization. The lattice
constant is $a$.} \label{figmodel}
\end{figure}

The single-particle eigenstates   
$|\alpha e\rangle$ and  
$|\beta h\rangle$ with the respective eigenvalues   
$\epsilon_{\alpha}^e$ and $\epsilon_{\beta}^h$ are obtained by diagonalizing  
the tight-binding Hamiltonian in the site-representation   
\begin{equation}      
  \widehat{H}_{0} = \sum_i\eta_i^e \widehat e^{\dag}_i  
\widehat e_i  
+J^e\sum_{<ij>}\widehat e^{\dag}_i \widehat e_j  
+\sum_i\eta_i^h \widehat h^{\dag}_i \widehat h_i+  
J^h\sum_{<ij>} \widehat h^{\dag}_i \widehat h_j  
\label{eq:H0} 
\end{equation}   
which contains the energies $\eta_i^{e,h}$ of the isolated sites 
contributing to the conduction and valence bands, respectively, and  
the nearest-neighbor couplings $J^{e,h}$ in these bands.   
The system consists of $N$ sites, and periodic boundary conditions 
are applied. In Eq.~(\ref{eq:H0}), the notation $<ij>$ indicates that 
the sums are 
limited to the nearest neighbors. For an ordered semiconductor
structure, all $\eta_i^{e}$ and all $\eta_i^{h}$ are taken to be identical.
For the description of disorder, the site energies are randomly varied 
within a prescribed range. We consider only diagonal disorder, i.e., the
coupling strength $J^{e,h}$ remains unchanged under all conditions.
A pictorial view of this tight-binding model is   
shown in Fig.~\ref{figmodel}.   

At this point a remark is on order. The sites of our tight-binding
model do not represent atoms. They rather represent building blocks which
allow us to model the band structure of the
ordered system
close to the band extrema
defining the gap with reasonable effective masses by choosing
the nearest-neighbor couplings $J^{e,h}$ and their nearest-neighbor
separation $a$ appropriately (see next
subsection). 

In the site-representation, the many-body Coulomb interaction reads  
\begin{equation}      
  \widehat H_{\mrm{Cb}} =   
  \frac12 \sum_{i,j} \sum_{\lambda}    
  V_{ij} \,  
 \widehat a^{\dag}_{\lambda i} \widehat a^{\dag}_{\lambda j}   
 \widehat a_{\lambda j} \widehat a_{\lambda i}  
-  
  \sum_{i,j}    
  V_{ij} \,  
 \widehat e^{\dag}_{i} \widehat h^{\dag}_{j}   
 \widehat h_{j} \widehat e_{i},  
\label{ham_Coul}
\end{equation}   
where only monopole-monopole terms have been included. 
In Eq.~(\ref{ham_Coul}), the first  
sum with $\lambda = e,h$ and $\widehat a_{ei}^{\dag}=\widehat e_i^{\dag}$,  
$\widehat a_{eh}^{\dag}=\widehat h_i^{\dag}$, etc., represents 
the repulsive  
Coulomb interaction for carriers within one band  
and the second term  
describes the attractive Coulomb interaction between electrons and  
holes. The regularized Coulomb matrix element considered here is given by 
\begin{equation} 
  V_{ij} = \frac{U_0}{|i-j|a + a_0} ,
\label{eq:Coulomb} 
\end{equation} 
with a positive constant $U_0$.  
Here, $a$ is the 
nearest-neighbor
site-separation and the term $a_0=0.5\,a$ removes the   
unphysical singularity of the lowest excitonic bound state arising   
from the restriction to monopole-monopole terms in one dimension.
In a one-dimensional semiconductor quantum wire it takes into account the
finite cross section of the structure \cite{banyai}.
  
In the single-particle eigenbasis, the interaction  
Hamiltonian reads for our two-band model  
\begin{equation} 
  \widehat H_{\mrm{Cb}} =   
\frac12 \sum_{\alpha, \beta} \sum_{\alpha', \beta'}    
\left[  
  \sum_{\lambda}  
  V_{\alpha, \beta}^{\alpha', \beta'} (\lambda,\lambda)  
  \widehat a^{\dag}_{\lambda \alpha'} \widehat a^{\dag}_{\lambda \beta'}   
  \widehat a_{\lambda \beta}\widehat a_{\lambda \alpha}  
- 2  
  V_{\alpha, \beta}^{\alpha', \beta'} (e,h)  
  \widehat e^{\dag}_{\alpha'} \widehat h^{\dag}_{\beta'}   
  \widehat h_{\beta}\widehat e_{\alpha}  
\right],  
\end{equation}
where  
\begin{equation}
V_{\alpha, \beta}^{\alpha', \beta'}(\lambda,\lambda')=  
\sum_{i,j}\psi^{\lambda*}_{\alpha'}(i)\psi^{\lambda'*}_{\beta'}(j)  
V_{ij}\psi^{\lambda'}_{\beta}(j)\psi^{\lambda}_{\alpha}(i)  
\end{equation}
and the $\psi^{\lambda}_{\alpha}(i)=\langle i|\alpha \lambda\rangle$   
are the single-particle eigenstates of $\widehat H_0$ in site representation.  
As we are interested in the low-density regime, the  
repulsive part of the interaction is omitted
in the remainder of this paper.  
  
For the calculation of optical absorption and luminescence spectra, we
also need the Hamiltonian contributions related to the quantized light field 
and the light-matter interaction. They are given by
\begin{equation}
  \widehat H_{\mrm{EM}} =  \widehat H_{\mrm{F}} + \widehat H_{\mrm{D}}
\end{equation}
with the free field Hamiltonian
\begin{equation}
\widehat H_{\mrm{F}} = \sum_q \hbar \omega_q \pa{ \widehat b^{\dag}_q 
\widehat b_q + \frac12 }
\end{equation}
and the light-matter interaction in dipole approximation,
\begin{equation}
  \widehat H_{\mrm{D}}    
  = -\sum_{\alpha \beta} \sum_q i {\mathcal E}_q \mu_{\alpha \beta}   
\widehat e^{\dag}_{\alpha} \widehat h^{\dag}_{\beta} \widehat b_q +   
  \sum_{\alpha \beta} \sum_q i {\mathcal E}_q \mu^*_{\alpha \beta}   
\widehat h_{\beta} \widehat e_{\alpha}\widehat b^{\dag}_q.
\end{equation} 
Here, the photon creation and annihilation operators are denoted by 
$\widehat b^{\dag}_q$ and $\widehat b_q$, respectively, ${\mathcal E}_q  
=\sqrt{\hbar \omega_q /2\epsilon_0}$ 
is the vacuum-field amplitude \cite{PQE} and $\mu_{\alpha \beta}$  
denotes the optical matrix element between single-particle  
eigenstates. The latter is obtained by transforming the optical  
matrix elements in site representation, $\mu_{ij}$, into the  
the basis of the single-particle eigenstates. We take $\mu_{ij}=  
\mu_0 \delta_{ij}$ for simplicity.
Note that within this approach it is not necessary to
consider radiative excitonic vs. plasma recombination separately.
Both processes are equally well contained in the total Hamiltonian.
For details see \cite{Kira98,PQE,Kira01,Chatterjee04,Hoyer03,Hoyer04}.

\subsection{The underlying ordered system}  
\label{ordered}  
 
\begin{figure}[htb!]\centering   
\resizebox{10cm}{!}{\includegraphics{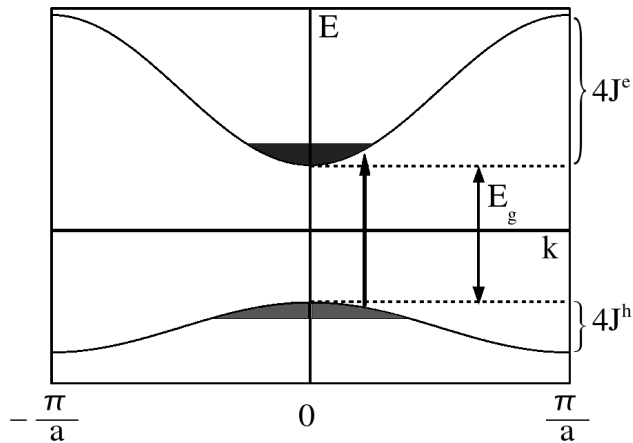}}   
\caption{\small Band structure of the ordered system  
representing a direct semiconductor.   
The shaded areas indicate occupied states due to pumping in the  
conduction and valence bands. The arrow indicates  
an allowed optical transition in a single-particle  
picture. $E_g$ denotes the band gap in the ordered case.} \label{figdirect}   
\end{figure}

\begin{figure}[htb!]\centering   
\resizebox{10cm}{!}{\includegraphics{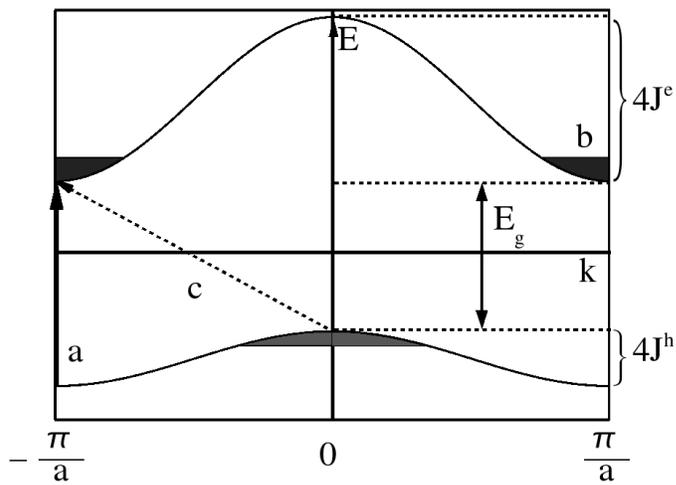}}   
\caption{\small 
Band structure of the ordered system representing an indirect
semiconductor. The shaded areas indicate occupied states due to
pumping in the conduction and valence bands. The solid arrow indicates
the lowest allowed optical transition connecting the minimum of the
valence band (a) with the minimum of the conduction band (b) in the
single-particle picture. The dashed arrow denotes an indirect
transition which is forbidden in the ordered case. It becomes partly
allowed if disorder is introduced. $E_g$ denotes the band gap in the
ordered case. 
} \label{figindirect}   
\end{figure}

The ordered system is described by $\eta_i^e=\eta_0/2$ and 
$\eta_i^h=\eta_0/2$.
By choosing the sign of the coupling matrix element according to either
$\mrm{sgn}(J^e)=\mrm{sgn}(J^h)$ or $\mrm{sgn}(J^e)=-\mrm{sgn}(J^h)$, one
can describe a direct or indirect semiconductor, respectively; see 
Figs.~\ref{figdirect} and \ref{figindirect}. Note that we are  
applying the electron-hole picture. The parameters are chosen
such that realistic effective masses are obtained
while a numerical treatment remains feasible. The system  
has a length $L=Na$ and periodic boundary conditions are applied.
In a tight-binding model the effective mass is given by  
\begin{equation}
m_{\mrm{eff}}^{\lambda}   
= \frac{\hbar^2 N^2} {2  \ab{J^{\lambda}} L^2}  
= \frac{\hbar^2} {2  \ab{J^{\lambda}} a^2}.  
\end{equation}
  
Since each site supports a single state in the  
valence band and a single state   
in the conduction band, the number of eigenstates  
and thus the number of optical transitions is determined by N.   
The numerical computations are limited to a maximum number of sites  
$N_{\mrm{max}}$. Therefore, our calculated spectra 
correspond to a series of discrete transitions  
in the ordered system instead of a true continuum.   
The combination $|J^{\lambda}|L^2/N^2$    
determines the effective masses in the band $\lambda$. Calculations  
are only meaningful if their results do not critically  
depend on $N$ for fixed $|J^{\lambda}|L^2/N^2$.  
This condition is fulfilled for all choices of the parameters  
used in this paper: $N=10$ to $N=30$, $a = 5$ nm,   
$\ab{J^e} = 8$ meV, $\ab{J^h} = 8/3$ meV, leading to effective masses  
\begin{equation}  
\frac{m_{\mrm{eff}}^{e}}{m_e} = 0.17, \;\;\;\; \;\;\;\; 
\frac{m_{\mrm{eff}}^{h}}{m_e} = 0.51,  
\end{equation}
where $m_e$ is the free electron mass.  
In a realistic semiconductor, the masses determine  
the (three-dimensional) exciton binding energy  
\begin{equation}
E_{\mrm{B}} = \frac{e^4 m_r}{2 \epsilon_{\mrm{bg}} \hbar^2}  
\end{equation}
with the background permittivity $\epsilon_{\mrm{bg}}$ 
and the reduced mass given by
\begin{equation}
\frac{1}{m_r} = \frac{1}{m_{\mrm{eff}}^e} + \frac{1}{m_{\mrm{eff}}^h}.
\end{equation}
In our model the numerical value of the binding energy is adjusted to be 
close to $E_{\mrm{B}}\simeq 10$\,meV by choosing an appropriate factor $U_0$ 
in $V_{ij}$.

Again we would like to point out that the values chosen for the
effective masses and the exciton binding energy characterizing the
many-particle interaction do not refer to any specific semiconducor
structure but rather serve as pure model parameters in this
case study.

\subsection{The disordered system}  
\label{disordered}  

The disordered system is modeled by assigning random values drawn from
a box distribution of width $W^e$ and $W^h$ to the site energies $\eta_i^e$ 
and $\eta_i^h$, respectively\cite{boxd}. In the present paper,
only uncorrelated disorder is modeled by choosing electron and hole
energies independently of one another. In principle, however, also
correlated or anticorrelated disorder could be used, where the upper and
lower energies of one site deviate in opposite or equal directions from
their average values on an absolute energy scale. 
 
We scale the disorder energies with the reciprocal masses according to
\begin{equation}
\frac{W^e}{W^h}=\frac{|J^e|}{|J^h|}
\end{equation}
in order to model disorder dueto fluctuations of confinement in
quantum-confined heterostructures. The dimensionless parameter  
\begin{equation}
\frac{W}{J}=\frac{W^{\lambda}}{J^{\lambda}}
\end{equation}
characterizes the strength of the disorder. In the following, we
use the term of a ``direct-based model'' (``indirect-based model'')
if the underlying ordered semiconductor is a direct (indirect) one, i.e.,
depending on whether $J^e$ and $J^h$ have equal or opposite sign.

In some figures, we show configurationally averaged data in order to 
smooth out the mesoscopic fluctuations due to small system size. This 
is achieved by drawing $M$ different realizations of   
disorder energies from a given distribution and
averaging over the calculated spectra. We also study  
the full distribution of certain features instead of
averages.   
    
\section{Microscopic theory for absorption and photoluminescence}  
\label{approach}  
\subsection{Absorption}  
The linear optical absorption can be calculated in a semiclassical 
framework from the optical polarization  
\begin{equation}
 P = \qav{\widehat P}=\sum_{\alpha   
\beta}(\mu_{\alpha \beta}^*  
P_{\alpha \beta}+\mu_{\alpha \beta}P_{\alpha \beta}^*),
\end{equation} 
with  
\begin{equation}
{P_{\alpha \beta} = \qav{\hat p_{\alpha \beta}}=  
\qav{\hat h_{\beta} \hat e_{\alpha}}.}
\end{equation}  
The interband coherence $P_{\alpha\beta}$ is obtained by
solving the semi\-con\-duc\-tor Bloch equation \cite{schwarz}
{\begin{eqnarray}
  i \hbar \frac{\partial}{\partial t} P_{\alpha \beta}   
  &=& \pa{\epsilon^e_{\alpha} + \epsilon^h_{\beta} -i \gamma}   
P_{\alpha \beta}   
  - \sum_{\alpha' \beta'}   
V_{\alpha' \beta'}^{\alpha \beta} P_{\alpha' \beta'}  
  -  \mu_{\alpha \beta} E(t),  
\label{eq:sbe}  
\end{eqnarray}}
written in the single-particle eigenbasis, where $\gamma  
=\hbar/{\tau}$ is a phenomenological dephasing term. 
Here, we have   
used the low-intensity limit, i.e., $P$ is linear in the field $E(t)$,
and thus neglected all intraband  
densities of the form $\ev{a^{\dagger}_{\lambda,\alpha} a_{\lambda,\alpha'}}$.  
Furthermore, we have not included microscopic Coulomb scattering, but  
used the constant damping instead.

Equation~(\ref{eq:sbe}) 
can be solved analytically via Fourier transformation.  
The solution is obtained most easily by introducing the exciton pair  
basis $\Phi^{\nu}$ given by the solution of the source-free (homogeneous)
part of   
Eq.~(\ref{eq:sbe})
\begin{equation}
\sum_{\alpha',\beta'}  
M_{\alpha \beta}^{\alpha' \beta'} \Phi^{\nu}_{\alpha' \beta'}   
  = \epsilon_{\nu} \Phi^{\nu}_{\alpha \beta} , \label{eom}  
\label{eq:wannier}  
\end{equation}
with  
\begin{equation}     
M_{\alpha \beta}^{\alpha' \beta'} =   
\delta_{\alpha \alpha'}  
  \delta_{\beta \beta'} \pa{\epsilon^e_{\alpha} +   
\epsilon^h_{\beta}}  
  -  V_{\alpha' \beta'}^{\alpha \beta}.  
\end{equation}   
Note, that we do not apply a factorization into relative and  
center-of-mass wave functions. Instead, $\Phi^{\nu}$ and  
$\epsilon_{\nu}$ denote the $\nu^{\mrm{th}}$    
two-particle eigenstate and eigenenergy, respectively, 
i.e.,   
\begin{equation}{\Phi_{\alpha \beta}^{\nu}=  
(\langle \alpha e | \otimes \langle \beta h| ) | \Phi^{\nu} \rangle}\end{equation}     
is the two-particle eigenstate projected onto the single-particle  
pair-state basis in the Hilbert space  
$\mathcal{H}^e \otimes \mathcal{H}^h$.  
  
Using Eq.~(\ref{eq:wannier}), we can analytically  
invert the Fourier transform of Eq.~(\ref{eq:sbe}) and obtain  
\begin{equation}{  
\chi''(\omega)=\sum_{\nu}\frac{|\mu_{\nu}|^2}{\epsilon_{\rm bg}}  
\frac{\gamma}{(\epsilon_{\nu} -\hbar \omega)^2+\gamma^2}  
\label{wichtig1}}\end{equation} 
for the imaginary part $\chi''(\omega)$ of the linear susceptibility  
$\chi(\omega) = P(\omega)/E(\omega)$. This quantity is directly 
proportional to the linear optical absorption spectrum. 
In Eq.~(\ref{wichtig1}),  
$\mu_{\nu}$ is the optical matrix element in the two-particle basis, i.e.,   
\begin{equation}{  
\mu_{\nu} = \sum_{\alpha \beta} \mu_{\alpha \beta}   
\Phi^{\nu}_{\alpha \beta}.
}\end{equation}
  
\subsection{Photoluminescence}  
%
We exclusively consider steady-state photoluminescence which is a good approximation
for sufficiently slowly varying carrier distribution. In that case, a good measure for 
the photoluminescence spectrum is given by
%
\begin{equation}
I_{\mrm{PL}}\pa{\omega_q} 
= 
\frac{\partial}{\partial t} \qav{\widehat b^{\dagger}_q \widehat b_q},
\end{equation}  
%
the rate of emitted photons\cite{PQE}. In principle, direction and magnitude of $\vec{q}$ determine
emission direction and frequency; here, we limit ourselves to only one specific direction
such that the scalar wave number $q$ determines the frequency $\omega_q = c q$.
Following a similar derivation as given in 
Refs.~\cite{Hoyer03,Hoyer04} 
for ordered semiconductor
heterostructures, this quantity can be expressed as  
\begin{equation}{  
I_{\mrm{PL}}\pa{\omega_q} =  \omega_q \sum_{\nu}   
  \frac{|\mu_{\nu}|^2  }{\epsilon_{\rm bg}} 
 \frac{ \gamma\, S_\nu}  
  {\pa{\hbar \omega_q - \epsilon_{\nu}}^2 + \gamma^2}  
\label{wichtig2}  
}\end{equation}
even for the disordered case. This formula is in close analogy to 
Eq.~(\ref{wichtig1})  
and exhibits the same resonances as can be seen from the energy 
denominator. As in  
Eq.~(\ref{wichtig1}), we have used the low-density limit in which we 
have not taken  
into account the microscopic Coulomb scattering which would lead to a 
state- and  
frequency-dependent $\gamma$.\cite{Chatterjee04} Thus, in reality the 
lineshape  
of the resonances is not a simple Lorentzian as is considered here.
However, for sufficiently low 
densities   
and for the case of strong disorder, we expect the result to be dominated by  
the inhomogeneous broadening such that the details of the homogeneous 
line shape  
are less important than for ordered semiconductors.   
  
In general, the source term $S_\nu$ contains true two-particle correlations 
of  
the form $\qav{\widehat e^{\dagger}\widehat h^{\dagger} \widehat h 
\widehat e}$.  
These correlations can either describe a Coulomb-correlated plasma 
beyond the pure  
Hartree-Fock limit\cite{Hoyer02c} or additional bound excitons 
\cite{Chatterjee04}.  
In the absence of any exciton formation, we find that these four-point  
quantities in the limit of a correlated plasma are given by 
\cite{Hoyer03,Hoyer04}  
\begin{equation}{  
S_\nu  
=  
\sum\limits_{\alpha \beta} \left| \Phi^{\nu}_{\alpha \beta} \right|^2
      f^e_{\beta} f^h_{\alpha}. \label{slam}  
}\end{equation}
Note that this source term does not result from a Hartree-Fock
factorization of the underlying correlation function
\cite{Hoyer03,Hoyer04}. Such an approach would yield a term
which is non-diagonal in the indices $\nu$ and $\nu'$ characterizing the
many-particle states and can lead to 
unphysical negative luminescence intensity. Only by considering in addition
the correlations beyond Hartree-Fock these non-diagonal terms cancel and
only one $\nu$-sum over pair states remains in Eq.~(\ref{slam}). In general,
the diagonal source term can contain contributions from both plasma and 
correlated bound excitons. In our situation of not too low temperatures it
is known \cite{Hoyer03} that the latter can safely be neglected.
Thus, the source term given in Eq.~(\ref{slam})
describes the emission of a correlated electron-hole  
plasma and the formation of incoherent bound exciton populations is not taken
into account. 

A more complete theory should be based on a fully dynamic  
description including the relaxation on a microscopic level.  
As stated above, it is presently not feasible for us to numerically implement this approach
for a strongly disordered interacting system. Thus, we assume stationary  
quasi-thermal distributions for the single-particle distributions  
$f^{e,h}_{\alpha}  = f^{e,h}_{\alpha \alpha}$ corresponding to the   
diagonal elements of the intra-band quantities   
$f^h_{\alpha \beta} = \qav{ \hat h^{\dag}_{\alpha} \hat h_{\beta} }$   
and $f^e_{\alpha \beta} = \qav{ \hat e^{\dag}_{\alpha} \hat e_{\beta} }$.  
The distributions are chosen such that a given density of carriers   
$\rho=10^{-3}$ per site is achieved.  
  
  
\section{Results and discussion}  
\label{results}   
 
In the following, we analyze 
disorder-induced effects on semiconductor 
absorption and luminescence in the limit of small
carrier densities. 
We concentrate, in particular, 
on situations where optically active exciton populations 
are not expected to play a crucial role.
More specifically, we 
investigate both the direct- and indirect-based models 
for different  
carrier temperatures and various amounts of disorder 
by explicitly evaluating 
Eqs.~(\ref{wichtig1}) and~(\ref{wichtig2}).
 
\subsection{Absorption and photoluminescence spectra}  
 
\begin{figure}[htb!]\centering   
{\resizebox{12cm}{!}{\includegraphics{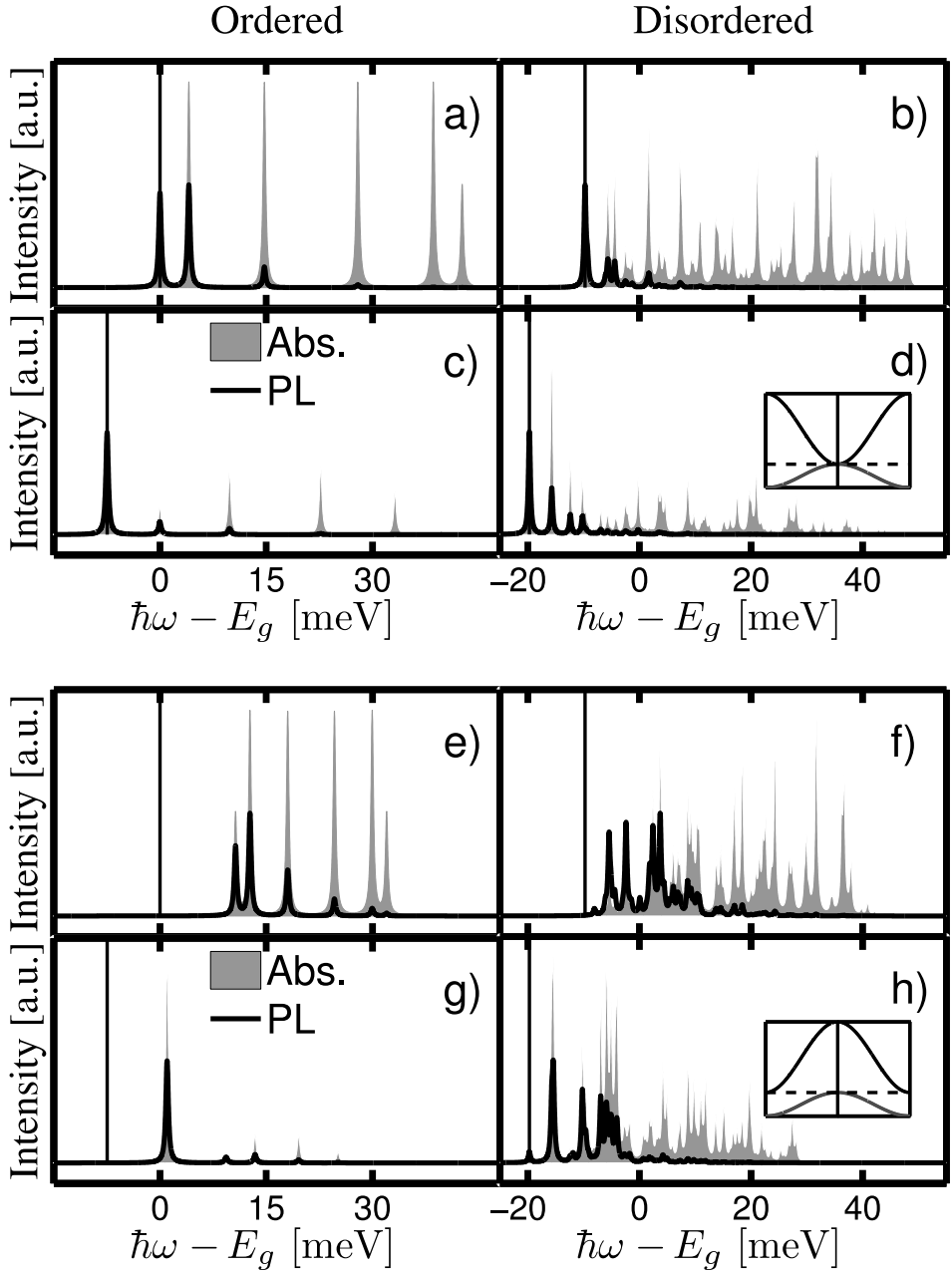}}}
\caption{\small Optical absorption (shaded areas) and luminescence  
(solid lines) spectra for ordered and disordered   
non-interacting and interacting direct- (upper panel) and 
indirect- (lower panel) based models.
Insets give the band structure of these two ordered
cases, showing that the
zero of
the photon energy has been chosen to coincide with the minimal
state-energy separation, see Figs.~
\ref{figdirect} and \ref{figindirect}. The thin vertical line
indicates the smallest energy separation for any given case.    
a), e) 
Absorption and luminescence spectra for an ordered non-interacting model.
The quasi-discrete spectra result from the system being composed from
only N = 10 sites.  
b), f)
The same situation, but uncorrelated disorder is
introduced ($W = 4J$).  c), g) Spectra of the ordered interacting model.  
d), h) Spectra of the disordered interacting model. 
Spectra for a single realization of disorder are
shown. 
Note that due to the k-selection rule the optical transitions
start at higher energies than
the minimal energy separation in the ordered 
indirect-based models, e) and g), 
while disorder tends to close this gap, f) and h).
} \label{figorddis}   
\end{figure}

For the 
direct- and 
indirect-based model, absorption and luminescence spectra are   
shown in Fig.~\ref{figorddis} for a system of $N=10$  
sites. The absorption spectrum of the ordered interaction-free case, 
Fig.~\ref{figorddis}a,   
shows 
four doubly degenerate transitions and two nondegenerate ones  
at the edges of the spectrum. The transitions at the
energetically lowest interband energy difference
(shown as a vertical line in Fig.~\ref{figorddis}
and taken as the zero of the energy scale
)
is indirect in k-space 
for the indirect-based model
(indicated by a ``c'' in Fig.~\ref{figindirect})
and thus forbidden.
In the ordered case, the only possible direct transitions connect 
states that are  
either at the top of the conduction band or at the bottom of the  
valence band. Since the corresponding carrier occupation is exponentially 
small,  
the luminescence signal is small as well (not seen in  
the figure where normalized data are shown).  
As a general trend, we observe that the central position of the 
luminescence spectrum  is shifted towards lower energies compared to the  
symmetric absorption spectrum. This results from the   
carrier distributions which decreases with increasing electron and hole 
energy.  
Furthermore, the high-energy electron (high-energy hole) states in our present  
tight-binding model do not reflect states in a normal wide-band  
semiconductor. They might, however, be relevant for superlattices,  
polymers, or other narrow-band semiconductors.  
 
Disorder leads to a violation of the $k$-selection rule in  
optical transitions. Thus, transitions that are forbidden  
in the ordered case may become allowed in the presence of disorder. 
This particularly applies to transitions
between electron and hole states that form 
the lower and upper conduction and valence band extrema, respectively.   
This is seen in both the absorption  
and the luminescence spectrum, see Fig.~\ref{figorddis}b
for the indirect-based model.
Here, with disorder, 
the   
optical transitions extend  
down to photon energies corresponding to the smallest energetic separation  
across the gap. In the disordered situation, the red-shift of the center of  
mass of the luminescence spectrum relative to that of the absorption is much  
more pronounced than in the ordered situation.  
  
The same trend is observed also if the Coulomb interaction is considered, see
Fig.~\ref{figorddis}c-d. 
The difference with respect to the interaction-free case lies 
basically in the  
well-known fact that the interaction leads to an enhancement of the  
matrix elements of the energetically low transitions which is 
accompanied by  
a reduction for the high-energy ones. In the ordered case, a distinct   
excitonic resonance emerges while in the disordered  
case no single bound exciton can be identified, except perhaps   
for the very lowest transition close to -30 meV.  
For direct-based models, the situation is similar, except that the  
disorder-induced red shift of both spectra is less pronounced, since in this  
case, the energetically lowest interband 
transitions are allowed even in the ordered case.
  
\subsubsection{The Stokes-Shift}  

\begin{figure}[htb!]\centering   
\resizebox{10cm}{!}{\includegraphics{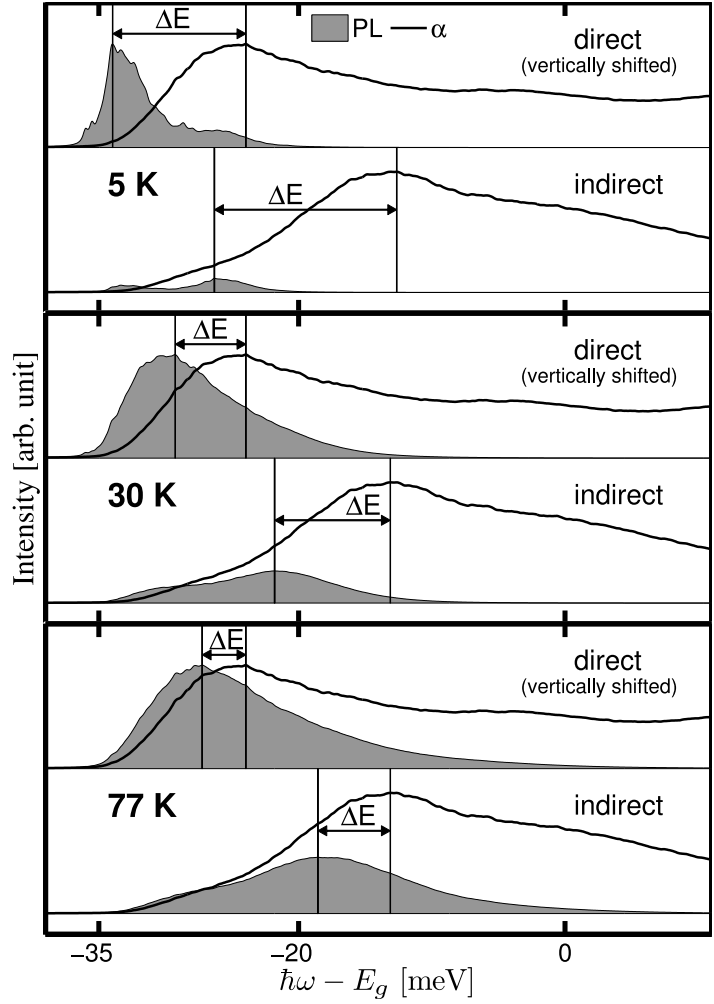}}
\caption{\small Spectra for disordered interacting systems, based  
on both direct and indirect models. Configurational averages are shown.  
$\Delta E$ indicated the Stokes shifts. The disorder parameter is  
$W=4J$. 
} \label{figdirindir}   
\end{figure}

The spectra of Fig.~\ref{figorddis} are 
computed for one particular realization  
of the disorder. The configurationally averaged spectra, see
Fig.~\ref{figdirindir}, 
more clearly show the red shift of the luminescence spectrum with respect  
to the absorption spectrum, i.e., the 
disorder-induced Stokes shift. To quantitatively analyze the Stokes shift, 
we determine the energetic red-shift of the maximum  
of the luminescence spectrum with respect to the maximum of the absorption and  
denote it by $\Delta E$. One clearly observes that $\Delta E$ is nonvanishing 
for both the direct-   
and indirect-based models; as shown in Fig.~\ref{figdirindir}, it is   
slightly larger in the second case.  
It is also seen that the Stokes shift increases
when lowering the temperature. 
A non-monotonous behavior at low
temperatures is obtained if phonon induced relaxation and hopping are
explicitly
included in the description \cite{wir}.
  
\begin{figure}[htb!]\centering   
\rotatebox{-90}{\resizebox{10cm}{!}{\includegraphics{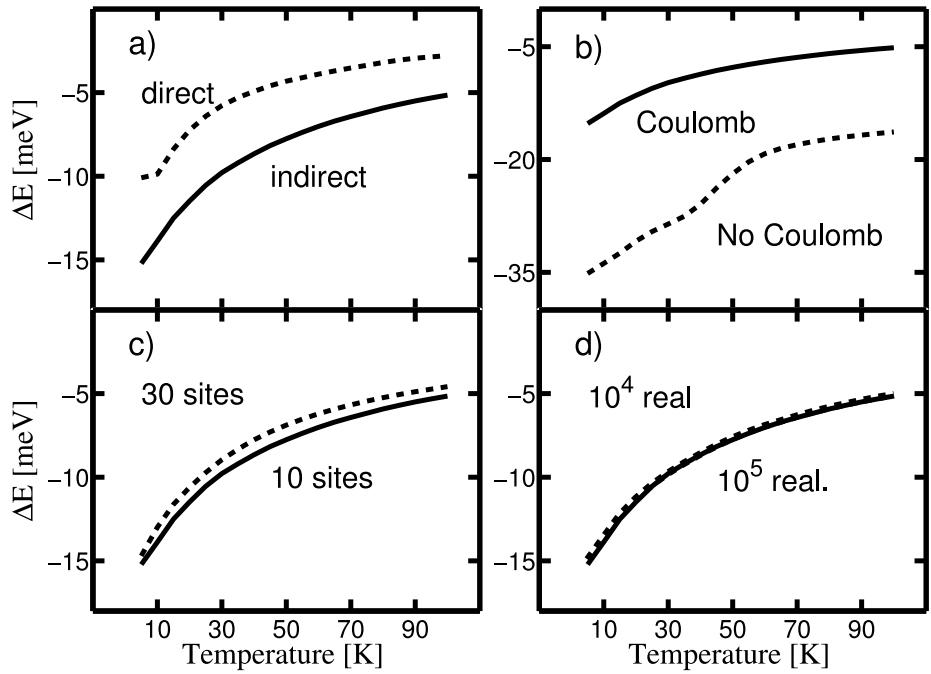}}}
\caption{\small Stokes shift for disordered models, $W=4J$,  
as function of temperature. a) Compares direct- and indirect-based  
models. b) Shows the influence of the Coulomb interaction and  
c) the influence of the number of sites $N$.  d) Demonstrates
convergence with respect to the number of realizations. b), c), and d)
correspond to the indirect-based model.
} \label{figtest}   
\end{figure}

\begin{figure}[htb!]\centering   
\rotatebox{-90}{\resizebox{10cm}{!}{\includegraphics{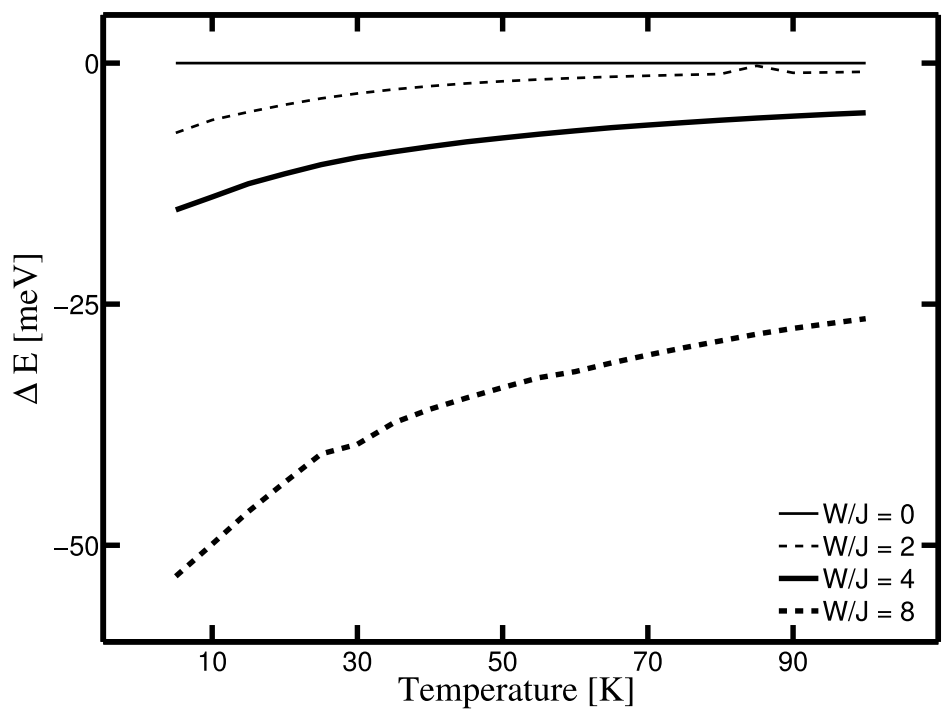}}}
\caption{\small The Stokes shift as function of temperature  
for different disorder parameters $W$. 
Every point corresponds to a Stokes shift obtained from the  
maxima of the configurationally averaged spectra.   
} \label{figconv}
\end{figure}

To complete the Stokes shift analysis, we present $\Delta E$ 
of the configurationally averaged spectra  
as function of temperature in Fig.~\ref{figtest}. Due to the quasi-thermal  
distribution of the carriers, the magnitude of the Stokes shift  
decreases with rising temperature. From Fig.~\ref{figtest}a, 
we see that this effect
is more pronounced for the indirect-based model. This is due to tailing of  
the spectra produced by the disorder-induced violation of the  
$k$-selection rule. These tails accommodate the low-energy carriers  
in both bands and lead to a larger weight at low photon energies  
in the luminescence spectra if compared to that of the direct-based model.  
The temperature dependence of the Stokes shift becomes more pronounced for 
elevated levels of the disorder. This is shown in Fig.~\ref{figconv} where  
$\Delta E$ is presented as functions of disorder and temperature 
for the indirect-based model. 
 
Figure~\ref{figtest}b shows that the interaction tends to reduce the  
Stokes shift. This can be explained by observing that the interaction  
leads to an enhancement of the optical matrix elements for  
low-energy transitions for both absorption and luminescence. Thus, even
in the absence of disorder, the absorption spectrum is 
energetically more
concentrated at low energies such that additional disorder is less
effective compared to the interaction-free case.
 
In order to judge whether the small number of  sites is a severe  
limitation for the present study, we compare  $\Delta E$ results of  
$N=10$ and  $N=30$ sites in Fig.~\ref{figtest}c.  
This indicates that $N=10$ sites provides fairly converged 
results since the Stokes shift only moderately depends on $N$. 
Fig.~\ref{figtest}d demonstrates the convergence of the results with respect  
to increasing number of realizations used in the configurational average.  
Once again, a good convergence is observed already with $10^4$ realizations, 
used in most of the computations. 
  
\subsubsection{Structure of the spectra}  
  
\begin{figure}[htb!]\centering   
\resizebox{10cm}{!}{\includegraphics{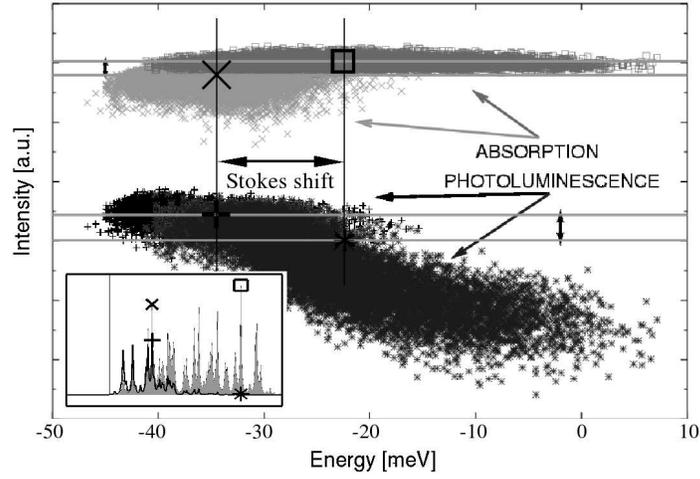}}   
\caption{\small Distribution of the maxima of the absorption and  
luminescence spectra for $10^4$ realizations. Here $T=10$K, $W=4J$.   
Squares: Distribution of the maxima of the absorption spectra.  
+: Distribution of  the maxima of the luminescence spectra.  
For every square the height of the corresponding  
luminescence spectrum at the  
same energy is given by a *. For every + the height of the   
corresponding absorption spectrum is given by an x. Big symbols   
denote center of mass of the corresponding cloud.   
} \label{figwolk1}   
\end{figure}

As seen in Fig.~\ref{figorddis}, the spectrum for a single  
realization consists of a very broad distribution of peaks of 
different height.  
In order to investigate to what extent
our procedure of configurational  
averaging is meaningful, we compare it to
an alternative approach. Therefore, we  
pick the spectral maxima for each configuration as the most relevant single   
piece of information and study their distribution.
In Fig.~\ref{figwolk1},  
the distributions of the luminescence and 
the absorption maxima are presented 
for $T=10$K and $W/J=4$ showing data  
for 10 000 disorder realizations.
The maxima of the absorption spectra are denoted by squares, while those  
of the luminescence spectra are given by $+$. The center of mass of  
each distribution is indicated by the corresponding large symbol.  

These averages of the maxima result in identical qualitative trends
for the Stokes shift as the maxima of the averages in Fig.~\ref{figorddis},
supporting the validity of the analysis of the spectra in terms of
configurational averages.

In addition to the two sets of distributions in Fig.~\ref{figwolk1},  
we also show the height of the absorption for every  
realization at that energy where the luminescence spectrum has its maximum  
(x), and vice-versa (*). Both types of data taken from luminescence
spectra (+ and x) show a decreasing trend with increasing energy. Since
this is not the case with data taken from absorption spectra (square
and *) this gives an additional justification for the definiton of the
disorder-induced Stokes shift. 

\begin{figure}[htb!]\centering   
\resizebox{10cm}{!}{\includegraphics{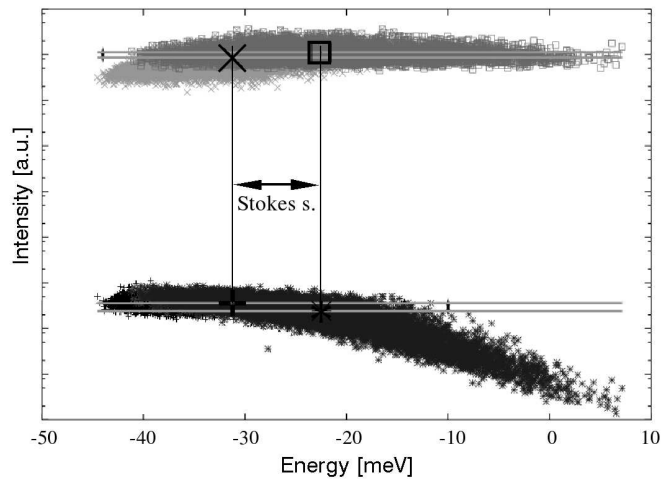}}   
\caption{\small Same as Fig \ref{figwolk1}, but $T=77$K.} \label{figwolk2}   
\end{figure}

In Fig.~\ref{figwolk2} the same distributions are shown, however, now at a  
higher temperature of $T=77$K. In this case the quasi-thermal  
distributions are more extended in energy, leading to a less   
pronounced dependence of the height of the maxima of the luminescence 
spectra on   
the emission energy. The magnitude of the Stokes shift is reduced in   
agreement with the data in Fig.~\ref{figconv}.  
  
\subsubsection{Disorder-induced convergence of spectra  
for direct- and indirect-based models}  
  
\begin{figure}[htb!]\centering   
\resizebox{10cm}{!}{\includegraphics{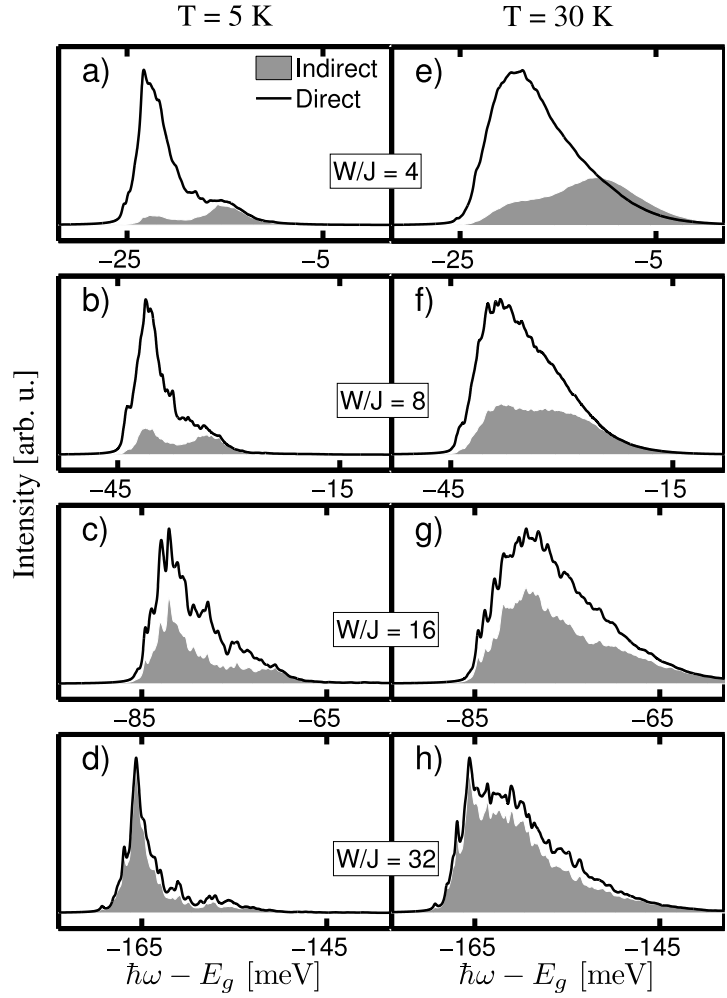}}   
\caption{\small Comparison of luminescence spectra for
direct and indirect based models for various disorder  
$W/J$ and for $T=5$K and $T=30$K.   
Filled areas: indirect based models. Full lines:
direct-based models. Configurationally averaged data are shown.  
} \label{figlifetime1}   
\end{figure}

Our computations show that the luminescence spectra for direct- and  
indirect-based models are different. This is most pronounced 
at low temperature, see   
Fig.~\ref{figlifetime1}, where averaged spectra for low ($T=5$K) and  
higher temperature ($T=30$K) are presented.   
Evidently, if disorder is increased, the spectra are  
expected to converge for both cases since band structure effects
are expected to become unimportant
for dominantly disordered systems. This trend is verified 
by Fig.~\ref{figlifetime1}. 
Nevertheless, we find it remarkable that even for relatively large  
disorder ($W/J=16$) the spectra still differ clearly from each other.  
This difference is more pronounced at low temperature.  
 
The luminescence spectra of Fig.~\ref{figlifetime1} show a double-peaked  
structure in the low-temperature data for the indirect-based models. 
To understand this feature, one has to realize that the spectra cover  
the total bandwidth of the tight-binding model.  
In particular, the double-peaked structure 
follows from: (i) transitions located  
close to the lowest energetic separation of electron and hole states   
and (ii) to transitions of type (a) and energetically higher ones  
indicated in Fig.~\ref{figindirect}, weighted by the carrier occupations.  
However, our tight-binding  
model describes real semiconductors adequately only in a small  
energetic window close to the fundamental gap. Energetically higher 
transitions   
have no direct correspondence to those in real   
semiconductors with their  
much larger band widths. Therefore, the double-peak structure seen   
at low temperatures should  
be taken as an artifact of the model, if one is interested in  
modeling a conventional wide-band 
semiconductor.  
On the other hand,   
this structure might correspond to realistic features for such  
systems that are  
characterized by small band widths. Examples include minibands in  
semiconductor superlattices or organic polymers, provided the  
underlying ordered system is an indirect semiconductor.  
    
\subsection{Thermodynamic Relation} 
%

Since the radiative recombination typically takes place on a nanosecond
time{\-}scale, it might be tempting to calculate the steady-state luminescence
spectrum under the assumption that the light-matter interaction acts
as a weak perturbation and to assume thermodynamic
quasi-equilibrium distributions for the photoexcited carrier
system, due to e.g.~fast phonon scattering. In this strict 
thermodynamic limit it is possible to derive a thermodynamic relation  
between photoluminescence and absorption \cite{Zimmermann88} 
\begin{equation} 
  I_{\rm TD-PL}(\omega) = g(\omega) \; \alpha(\omega) 
\label{eq:KMS_relation}, 
\end{equation} 
where $g(\omega)$ is given by a Bose-Einstein distribution.

For ordered semiconductors, however, it is well known and understood
that this approximation fails at the 1s-exciton resonance. Several 
recent experiments \cite{Chatterjee04,Schnabel92,Szczytko04} demonstrate
that the luminescence at the 1s-exciton resonance strongly violates  
Eq.~(\ref{eq:KMS_relation}) even after all excitation transients have 
been completely equilibrated. In this case, the nonthermal behavior 
originates from the strong depletion of optically active excitons  
implying that photoluminescence is not a weak perturbation for  
exciton distributions \cite{Kira01} and that, consequently,
the thermodynamic arguments cannot be used. Even though the single
particle distributions may be close to Fermi-Dirac distributions,
the two-particle excitonic distributions can be very nonthermal and
especially the 1s-distribution can exhibit considerable hole-burning.
Thus, the photoluminescence at the 1s-exciton resonance is in many 
situations dominated by correlated plasma contributions rather than
exciton populations. The result is a nonthermal 1s luminescence for
nearly disorderless systems \cite{Chatterjee04}.
However, it is also known that Eq.~(\ref{eq:KMS_relation}) typically
remains valid for photon energies corresponding to the continuum states
when the semiconductor carriers have reached a quasi-equilibrium after the
excitation. Thus, for elevated carrier densities and temperatures as they
occur e.g.~under lasing conditions when the Coulomb interaction is strongly 
screened and the excitonic resonance is bleached out, 
Eq.~(\ref{eq:KMS_relation}) can be safely used to calculate luminescence
from absorption spectra.

Our goal in the following section is to investigate whether the thermodynamic
relation can be recovered even for the correlated-plasma emission when the
system has a considerable amount of disorder.  Since we only study luminescence 
from systems with dilute carrier densities we can replace the Bose-Einstein
distribution $g(\omega)$ by a Boltzmann distribution 
$g(\omega) = N_g\, e^{-\hbar \omega/k_B T}$ with a suitable normalization 
constant $N_g$. 

To determine the nonthermal aspects of the luminescence spectrum,
we apply the following procedure: i) we first determine an effective
temperature $T_{\rm eff}$ by fitting 
$g(\omega) = N_g\, e^{-\hbar \omega/k_B T_{\rm eff}}$
to the continuum part of the ratio $I_{\rm PL}(\omega) / \alpha(\omega)$,
where Eq.~(\ref{eq:KMS_relation}) is known to be valid. ii) After this, we use
Eq.~(\ref{eq:KMS_relation}) to predict the thermodynamic value,
$I_{\rm TD-PL}(\omega_{1}) = g(\omega_{1}) \; \alpha(\omega_{1})$,
for the lowest excitonic resonance at the frequency $\omega_1$. 
iii) This allows us to quantify the degree of nonthermal behavior via a 
$\beta$ factor,
\begin{equation}
  \beta =
  \frac{I_{\rm PL}(\omega_{1})}{I_{\rm TD-PL}(\omega_{1})}
\label{eq:beta_factor},
\end{equation}
which has been determined according to Schnabel et al.~\cite{Schnabel92}.
In other words, the $\beta$ factor describes how strongly the calculated
luminescence $I_{\rm PL}(\omega_{1})$ is suppressed compared to the expected
thermodynamic value $I_{\rm TD-PL}(\omega_{1})$ as obtained from the absorption
spectrum via  Eq.~(\ref{eq:KMS_relation}). A strict thermodynamic value is
observed when $\beta = 1$ while $\beta \ll 1$ indicates a strong violation
of the thermodynamic relation.

\begin{figure}[htb!]\centering   
\rotatebox{-90}{\resizebox{10cm}{!}{\includegraphics{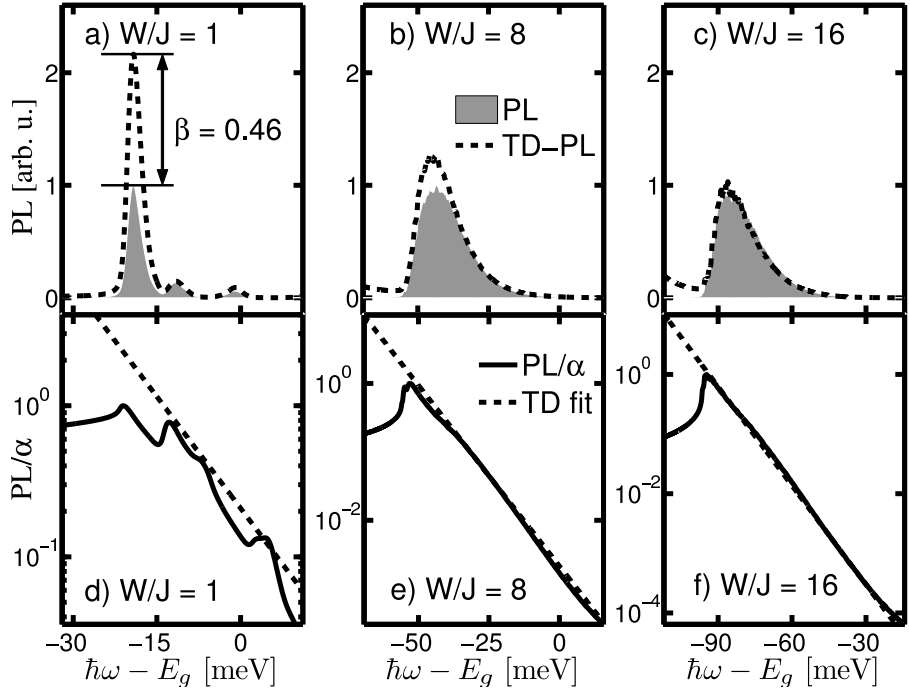}}}
\caption{\small 
Deviation of the luminescence spectra from the thermodynamical 
relation.
The calculated
luminescence spectra (filled area) are compared with the
luminescence derived from absorption $\alpha$
using the thermodynamic relation (dashed line) 
on a linear scale for $W/J=$1, 8, 16 in the upper row.
In the lower row the PL/$\alpha$ ratio (solid line) is 
fitted by the thermal line (dashed line) on a semi-log scale. 
} \label{fig1:KMS}
\end{figure}

This $\beta$ analysis reveals generic disorder dependent features which are
illustrated in Fig.~\ref{fig1:KMS} where the upper row
shows the actual luminescence spectra (filled areas) in comparison
with the thermodynamic spectrum (dashed line) constructed via
Eq.~(\ref{eq:KMS_relation}). The lower row compares the fitted Maxwell
distribution $g(\omega) = e^{-\hbar/k_B T_{\rm eff}}$ with the actual ratio
$I_{\rm PL}(\omega) / \alpha(\omega)$ for the corresponding spectra. These
computations were performed for a carrier temperature of 77\,K and three 
different
levels of disorder: weak ($W/J=1$, left column), intermediate ($W/J=8$, 
middle column),
and strong disorder ($W/J=16$, right column). All spectra are evaluated 
using 10 sites
and averaging over $10^4$ realizations is applied. For the nearly 
disorderless case 
($W/J=1$), we find $\beta = 0.46$  indicating a clear violation of the 
thermodynamic 
relation. When the disorder amplitude is increased, $\beta$ starts to approach 
unity. For the largest disorder used, we find $\beta = 0.93$, which shows that 
an increased amount of disorder helps to recover the thermodynamic 
relation even 
for the excitonic plasma luminescence. In this case, the actual and the 
thermodynamic luminescence spectra are very close to each other.

The role of disorder can be intuitively understood if we consider
that the excitonic luminescence peak consists of a quasi-continuous 
distribution of energies for the high disorder case while only a single
discrete excitonic energy is present for the low disorder case.
Since the strongly disordered emission peaks are weighted
thermodynamically in the total luminescence, a fully thermodynamic relation
is approached as the disorder increases.

\begin{figure}[htb!]\centering   
\resizebox{10cm}{!}{\includegraphics{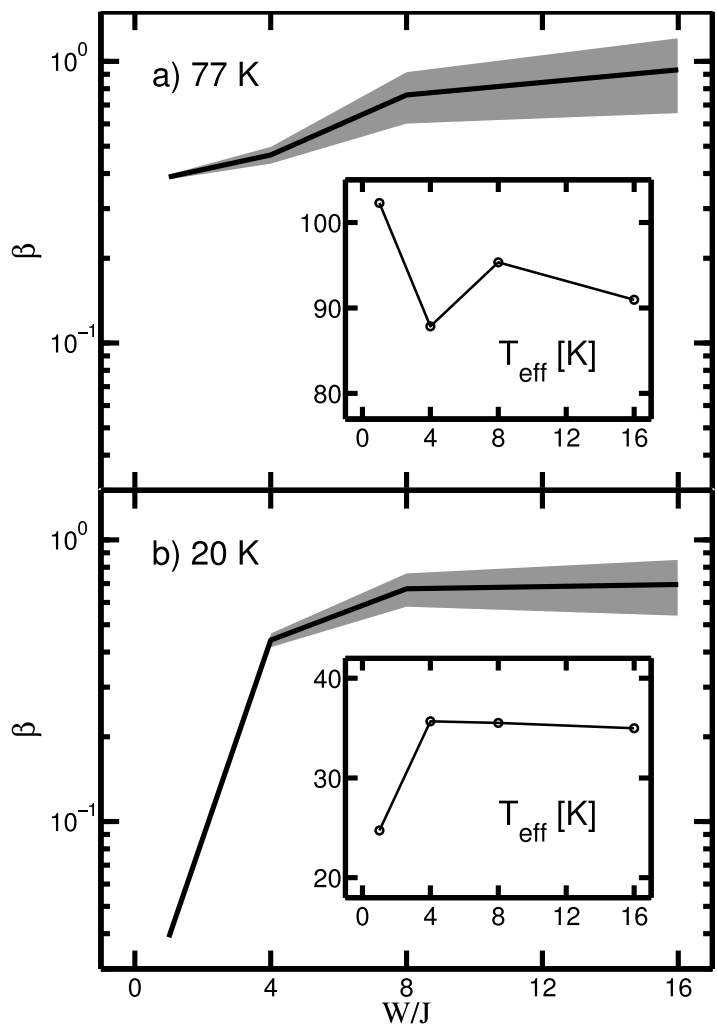}}   
\caption{\small 
The dependence of the amount of deviation from the
thermal value, $\beta$, as function of disorder (solid line). 
The error of the fit is also denoted (shaded area). Shown are data 
for $T=77$K (upper figure) and $T=20$K (lower figure). 
The insets show how $T_{\rm eff}$ changes with increasing disorder.
} \label{fig2:KMS}   
\end{figure}

Figure~\ref{fig2:KMS} shows $\beta$ (solid line) as a function 
of disorder for a lattice temperature of $T = 77$\,K (upper frame),
and $T = 20$\,K (lower frame); the inset displays the corresponding
$T_{\rm eff}$. We observe that for 20\,K carriers, the lowest $\beta$ lies
below 0.04 and increases monotonously with increasing disorder. Otherwise, 
similar qualitative behavior is observed as for 77\,K. For both 20\,K and 
77\,K, the disorder averaged spectra have noise. As a result, $\beta$ has 
error bars 
which increase for elevated level of disorder; the error is indicated by the 
shaded areas. The inset displays the effective temperature fitted from 
the luminescence tail. 

Actual experiments have only a limited access to the quantities determining 
the carrier system, e.g., luminescence and absorption spectra can be  
determined while exact carrier distributions remain usually unknown. 
The $\beta$-analysis procedure applied above only assumes that carrier
distributions are in quasi equilibrium such that one can conveniently 
use the same approach for an experimental analysis. As an additional theory  
feature, we can study how well the determined effective 
temperature corresponds to the actual carrier temperature. 
We observe that the effective temperature is always higher than the 
nominal carrier temperature used in the computations, which is
in agreement with previous findings for ordered semiconductors.

\subsection{The radiative lifetime distribution}  
  
If nonradiative recombination of electrons and holes can be 
neglected, 
the area under the stationary luminescence spectra directly reflects the   
radiative lifetime $\tau$ of the electron-hole system, 
i.e., the lifetime  
is simply inversely proportional to the area under the total  
spectrum. Thus, each spectrum for a given realization  
defines a single $\tau$ and the lifetimes   
calculated for many realizations define a
distribution of a certain shape.  
In practice, 
the lifetime can be calculated from Eq.~(\ref{wichtig2}) as   
$\tau^{-1}=\int d\omega\,I_{\mrm{PL}}(\omega)  $ yielding the relation  
\begin{equation}{
\frac{1}{\tau}\sim   
     \sum_{\nu} |\mu_{\nu}^2| S_{\nu} \varepsilon_{\nu}  
     \label{tau}.  
}\end{equation}
  
\begin{figure}[htb!]\centering   
\rotatebox{-90}{\resizebox{10cm}{!}{\includegraphics{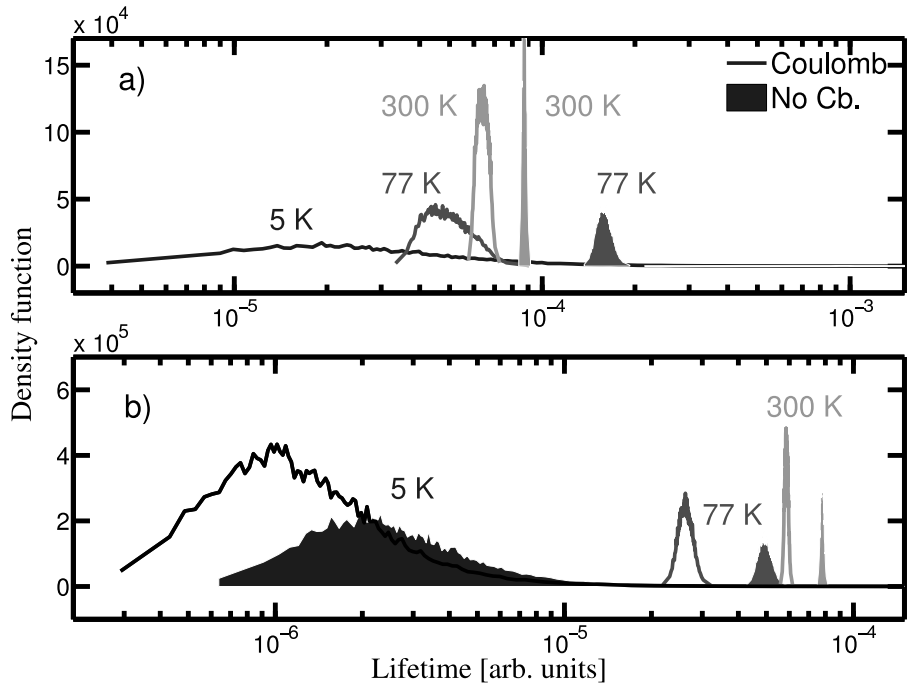}}}
\caption{\small Radiative lifetime distributions for   
an a) indirect-based model and b) the direct-based one
for various temperatures in the case of weak disorder, $W/J=1$. 
} \label{figlifetime2}   
\end{figure}

\begin{figure}[htb!]\centering   
\resizebox{10cm}{!}{\includegraphics{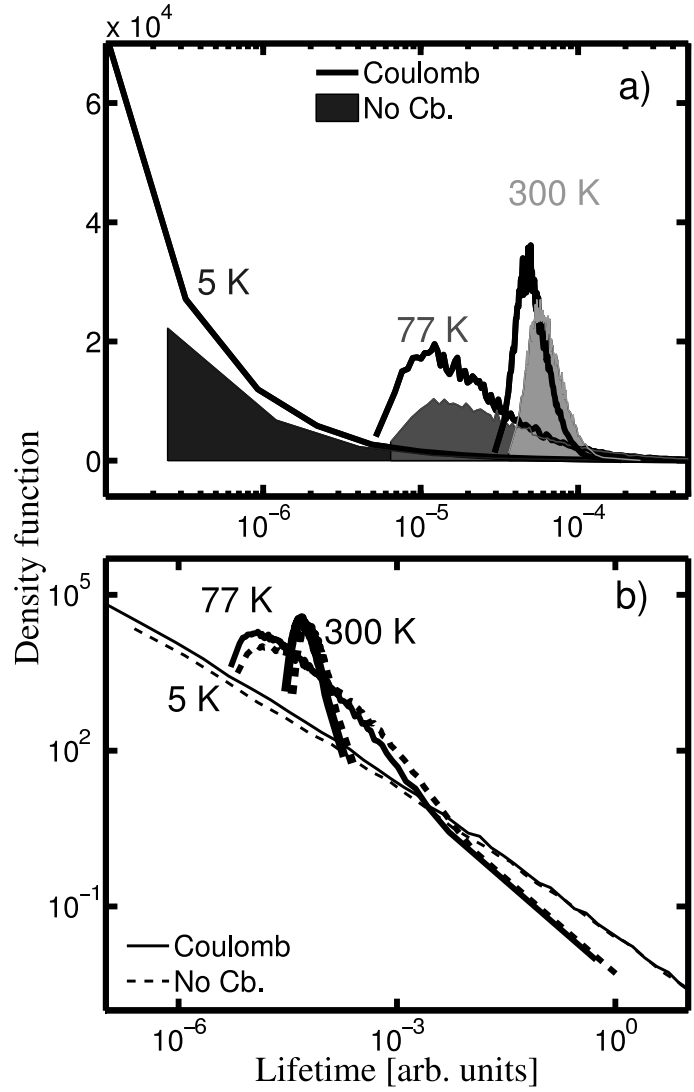}}   
\caption{\small Radiative lifetime distributions  
for an indirect-based model and for various temperatures for  
strong disorder, $W/J=16$, on a) semilog and 
b) double-logarithmical plots. 
} \label{figlifetime3}   
\end{figure}

For both the direct- and the indirect-based models we show such lifetime  
distributions for the non-interacting and the interacting case as function  
of temperature for a weakly and strongly disordered situation in   
Figs.~\ref{figlifetime2} and \ref{figlifetime3}.  
  
If the disorder is weak, i.e.,  
smaller than the exciton binding energy ($W/J=1$),  
the distributions for both non-interacting and interacting cases  
follow a log-normal form  
\begin{equation}{  
{\mathcal P}(\log \tau) \propto   
           \exp\left\{-\log (\tau/\tau_0)^2/{2\sigma^2}\right\}.  
}\end{equation}
For the indirect-based model (upper panel in Fig.~\ref{figlifetime2}),  
the distributions 
shift towards longer (shorter) times for the interacting  
(non-interacting) situation with increasing temperature. On the other hand,  
for the direct-based model both sets of distributions follow the same trend  
as function of temperature. The distributions become narrower on a logarithmic  
scale with increasing temperature.   
  
The interpretation is as follows: 
in the non-interacting case, the carriers occupy higher states  
for increasing temperature. In the indirect-based model,  
transitions between   
these states have larger matrix elements   
compared to those between the energetically low-lying states due to the  
still dominant indirect nature at weak disorder.   
Therefore, with increasing temperature the  
emission becomes increasingly more efficient, i.e., the lifetime  
decreases. This effect is overcompensated by 
the Coulomb interaction. In this case,  
already at low temperature, i.e., when low-energy transitions dominate, 
the interaction increases the otherwise small optical matrix elements
leading to an enhanced emission and short lifetimes. With increasing  
temperature energetically higher transitions become involved which are  
characterized by a smaller
Coulomb-induced enhancement. Thus, their weight decreases if compared to the  
low-temperature case. At high temperature, the distributions  
converge due to a smearing out of the distributions over a large energetic  
region and interaction is no longer important.  
  
For direct-based models, also for the non-interacting case
the low-energy transitions dominate the weight of the spectrum at low  
temperatures, thus both sets of distributions behave in a similar way.  
The temperature dependence of the interacting case is slightly more  
pronounced.  
 
Finally, at the end of this discussion of the lifetime we would like to
make a connection with different but related context. For relatively
large disorder ($W/J=16$), larger than the exciton binding energy,
both sets of distributions become quite similar to each other, see
Fig.~\ref{figlifetime3}. They both shift towards larger lifetimes with
increasing temperature. However, for low temperatures, they deviate
from a log-normal distribution. At the lowest temperature ($T=5$\,K),
only a monotonously decreasing ${\mathcal P}$ is observed as function of
increasing $\tau$. The explanation lies in the fact that there is a
cut-off at a minimum lifetime which is defined by the spatially direct
transition within a single isolated two-level system representing a
site. Its matrix element is simply given by the model parameter
$\mu_0$. Furthermore, the log-log plot presented in the lower panel in
Fig.~\ref{figlifetime3} shows that for strong disorder and low
temperatures the lifetime distribution follows  an overall power-law
dependence with an exponent close to unity. 

The above behavior of log-normal lifetime distribution for  
high temperatures and low disorder and power-law distribution  
for sufficiently low temperatures and strong disorder resembles  
the behavior seen in other disordered systems. 
In general terms, an electronic excitation 
may decay if the system is coupled to the outside, i.e., 
if it is an open system. This decay manifests itself in a 
homogeneous broadening of the sharp spectral line
of the electronic excitation in a closed system. In our system,
however, this homogeneous broadening is neither described 
by the Hamiltonian~\cite{gamma}, nor would it be possible to be seen, since
our spectrum is dominantly inhomogeneously broadened. Nevertheless, 
we are able to determine the radiative lifetime $\tau$
by integrating over the luminescence
spectrum, since we are considering a stationary 
situation. 

The lifetime is related to the coupling of the electronic
states that describe the excitation to the  
environment. So far, lifetime has been mostly
discussed in the context of electronic
transport \cite{mirlin,kottos,terraneo}. In that case,
it is the amplitude of the wave function at 
the position where the leads are attached to a system which determine the
lifetime. In our case, the lifetime is determined by the optical 
dipole matrix element $\mu_{\lambda}$.  From the Elliott formula, 
Eq.~(\ref{wichtig2}), we see that the lifetime distribution (or its 
inverse as recombination rate) follows the distribution of 
$S_{\nu}$, Eq.~(\ref{slam}), and through  
this quantity it is also based  dominantly on the wave function components. 
It is  known for the case of weakly disordered systems that the wave  
function components obey log-normal distribution  
in quasi-one dimensional systems which is applicable here.\cite{mirlin}  
Moreover, it has recently been shown \cite{kottos,titov} 
that resonance width  
distributions in diffusive systems show a crossover behavior 
between a log-normal and a power-law behavior. However, the exponent 
is different from the one obtained in the present calculation, but 
the discussion in this 
field are not yet settled. 
There are several works in the literature 
where power-law distributions of power equal to one are 
obtained.\cite{terraneo,new}

We would like to mention, that the appearance of log-normal behavior of 
the recombination times resembles also the hierarchical lifetime 
structure seen in spin--glass systems \cite{terraneo,casati,souletie}. These
consist of coupled two-level systems similarly to the disordered model of a 
semiconductor presented here. Presently, we do not see a link between 
these two fields but apparently an unusual analogy can be detected 
which deserves further investigation. 
 
The observation that the distributions for both the 
interacting and non-interacting cases show a similar behavior 
for large disorder demonstrates the fact that  
Coulomb interaction becomes less important with 
increasing disorder. Roughly speaking, for the short-range 
disorder potentials 
fluctuating on an energy scale on the order of the exciton binding energy
the dominant features of the luminescence spectra are determined by disorder. 
On the other hand, interaction-induced features will be clearly 
visible even for large disorder if the length scale 
of the disorder potential is larger than the exciton Bohr radius. 
This 
situation will be studied in a forthcoming paper. 

\section{Conclusions}   
\label{con}  

As a first step towards a fully microscopic theory of luminescence
in strongly disordered semiconductors, we have studied a
two-band tight-binding model including Coulomb interaction and diagonal
short-range disorder of arbitrary strength. Both absorption spectra and 
luminescence spectra have been calculated from modified Elliott formulas,
which include the electron-hole pair states in the presence of disorder and
the correlated-plasma source term for luminescence. The calculations are 
based on a direct diagonalization of the many-particle Hamiltonian.\cite{comp}
Certain simplifications have been made and discussed  
in order to obtain a computationally feasible description of 
the strongly disordered, interacting semiconductor. We believe that for the 
present case study these simplifications are justified. We have exclusively 
considered stationary luminescence and have ignored nonradiative processes.  

In its present state, our approach cannot directly be applied to 
those experimental data where the phonon-assisted dynamics 
is known to take place on a time scale longer than that of the
recombination. This is the case for low
temperatures, see, e.g., \cite{wir}. 
On the other hand, we have shown that
a description of radiative recombination in disordered
semiconductors is feasible treating
disorder and many-particle interactions on equal footing
without using special model assumptions
for bound-excitonic vs. plasma recombination.   

 In particular, we have in detail studied the influence of disorder, Coulomb
interaction, direct- or indirect-based models, and temperature on the
disorder-induced Stokes shift and the radiative lifetime distribution.
The results and their interpretation are applicable to stationary
luminescence measurements of strongly disordered semiconductors
including polymers~\cite{polymers} and superlatices~\cite{superlattice} for not
too low temperatures and for quasi-stationary situations where optically
active exciton populations are negligible. The cross-over from the low-temperature
to the high-temperature regime depends on details of the model, such as
density of states and degree of disorder. 
  
Time-resolved measurements or measurements on systems that are  
characterized by slow relaxation processes due to hopping events  
are outside the scope of this work, i.e., the low-temperature  
data of this calculation should be taken as a case study of our 
model system only. In realistic systems at extremely low temperatures,
hopping processes may become extremely slow such that the assumption of
quasi-thermal carrier distributions can no longer be justified. For 
higher temperatures, on the other hand, the observed trends can give 
support to an interpretation of stationary luminescence spectra of  
strongly disordered semiconductors.  



\newpage   
   

\begin{thebibliography}{99}
\bibitem{gamma}
More precisely, it is described with the phenomenological model parameter $\gamma$.
\bibitem{comp}
We have also calculated luminescence spectra dynamically 
using the equation-of-motion approach; 
the results (not shown here) give identical results, 
but with appreciably larger numerical effort.
\end{thebibliography}
\end{document}